\documentclass{article}

\usepackage{arxiv}

\usepackage[utf8]{inputenc} 
\usepackage[T1]{fontenc}    
\usepackage{hyperref}       
\usepackage{url}            
\usepackage{booktabs}       
\usepackage{amsfonts}       
\usepackage{nicefrac}       
\usepackage{microtype}      
\usepackage{lipsum}		
\usepackage{graphicx}
\usepackage{natbib}
\usepackage{doi}

\title{Coupling Efficiency and Laser-Induced Damage Threshold Characterization of an End-Capped Optical Fiber with a Sub-Nanosecond Pulsed Laser}

\author{ \href{https://orcid.org/0009-0006-9912-9016}{\includegraphics[scale=0.06]{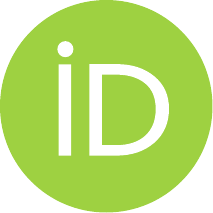}\hspace{1mm}Giorgos Stavrakakis}\\
	{Department of Chemistry, University of Crete}\\
    Foundation for Research and Technology Hellas\\
    Crete, Greece\\
	\texttt{gstav@iesl.forth.gr} \\
	\And
	\href{https://orcid.org/0000-0003-0648-4811}{\includegraphics[scale=0.06]{orcid.pdf}\hspace{1mm}Kostas G  Mavrakis}\\
	Foundation for Research and Technology Hellas\\ 
    Crete, Greece\\
	\texttt{mavrakis@iesl.forth.gr} \\ \\
    \And
	\href{https://orcid.org/0000-0002-6150-5593}{\includegraphics[scale=0.06]{orcid.pdf}\hspace{1mm}Mikis Mylonakis}\\
	Foundation for Research and Technology Hellas\\ 
    Crete, Greece\\
	\texttt{mylonakis@iesl.forth.gr} \\ \\
    \And
	\href{https://orcid.org/0000-0002-6438-0773}{\includegraphics[scale=0.06]{orcid.pdf}\hspace{1mm}Giannis  Zacharakis}\\
	Foundation for Research and Technology Hellas\\ 
     Crete, Greece\\
	\texttt{zahari@iesl.forth.gr} \\ \\
}

\renewcommand{\headeright}{Technical Report}
\renewcommand{\undertitle}{Technical Report}

\hypersetup{
pdftitle={Coupling Efficiency and Laser-Induced Damage Threshold Characterization of an End-Capped Optical Fiber with a Sub-Nanosecond Pulsed Laser},
pdfsubject={physics.optics},
pdfauthor={Giorgos Stavrakakis, Kostas G Mavrakis, Mikis Mylonakis, Giannis Zacharakis},
pdfkeywords={fiber endcaps, damage threshold, coupling efficiency},
}

\begin{document}
\maketitle
\setlength{\headheight}{23pt}

\begin{abstract}
	This paper presents experimental measurements of coupling efficiency and laser-induced damage thresholds for a polarization-maintaining (PM) fiber patchcord with integrated end caps, evaluated at 532 nm using a compact actively Q-switched DPSS laser with sub-nanosecond pulses. It also describes the development of a custom free-space coupling array through which the problem of low coupling efficiency was identified and successfully addressed. No instantaneous damage was observed at peak power densities exceeding 10 GW/cm$^2$. Sustained operation at 30 kHz was maintained over extended durations (>5 h) at peak power densities of $\sim$13 GW/cm$^2$, while prolonged 1 kHz operation led to gradual degradation at peak power densities of $\sim$24 GW/cm$^2$.
The broader context of this work is the investigation of stimulated Raman scattering (SRS) in optical fibers for nonlinear frequency conversion. This process requires the efficient delivery of high-peak-power pulses into the fiber, which serves as the nonlinear medium. In the course of these experiments, a substantial dataset was accumulated on fiber coupling performance and damage thresholds under repeated high-intensity illumination at 532 nm. These characterization data offer practical insight into the operational limits of end-capped PM fibers in demanding pulsed laser applications.

\end{abstract}


\section{Introduction}

Optical fibers offer significant advantages for laser beam delivery, including compact integration and mechanical stability\cite{hunter1996understanding}. In pulsed laser systems, performance is frequently constrained by coupling efficiency and laser-induced damage at the input facet. Sub-nanosecond pulses generate high instantaneous peak powers that can drive thermal absorption and nonlinear optical processes at the fiber facet \cite{abbasi2021highly}. Published data for end-capped PM fibers operated with sub-nanosecond pulsed lasers at 532 nm remain scarce, motivating the publication of measurements presented here. 

\section{Optical fibers}
The fiber patchcords under test, provided by SQS FiberOptics, were based on the PM 460HP polarization-maintaining fiber, with core, cladding and coating diameter of 3, 125 and 245 $\mu$m respectively, and terminated with FC/APC narrow key connectors on both ends. The patchcords featured integrated end caps of 400 $\mu$m length, with a measured polarization extinction ratio exceeding 20 dB. A fiber with total length of 20.00 m was used for Laser-Induced Damage Threshold measurements. A fiber of the same type with total length of 5.00 m was used for Coupling Efficiency measurements.

\section{Experimental setup}
An actively Q-switched DPSS laser emitting at 532 nm with sub-nanosecond pulse duration was used. Two repetition rates were investigated as summarized in Table 1. The average power at the input of the fiber was measured using a photodiode power meter calibrated for 532 nm. From the measured average power P$_{avg}$, and knowing the repetition rate f and pulse duration $\tau$ of the laser, the pulse energy and peak power were calculated as follows:

\begin{equation}
	E_{pulse}=\frac{P_{avg}}{f}
\end{equation}
\begin{equation}
	P_{peak}=\frac{E_{pulse}}{\tau}=\frac{P_{avg}}{\tau \cdot f}
\end{equation}

\begin{table}
	\caption{Laser Source and Operating Parameters}
	\centering
	\begin{tabular}{|l|l|l|}
		\toprule
		\textbf{Parameter}     & \textbf{1 kHz}     & \textbf{30 kHz} \\
		\midrule
		Wavelength & 532 nm  & 532 nm     \\
		Max average power & 720 $\mu m$ & 33 mW     \\
		Pulse duration & 544 ps       & 668 ps  \\
        Max peak power & 1323 W & 1647 W\\
		\bottomrule
	\end{tabular}
	\label{tab:table}
\end{table}

\begin{figure}[!ht]
    \includegraphics[]{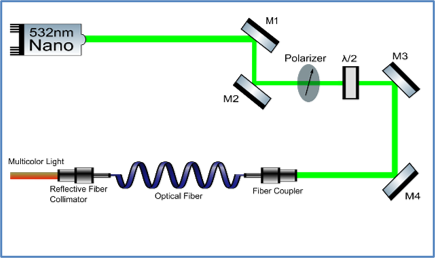}{}
	\centering
	\caption{Schematic of the free-space fiber coupling arrangement.}
	\label{fig1}
\end{figure}

Laser radiation was directed into the fiber via a fiber coupler (f=12 mm, NA=0.23, AR coated 390–670 nm, adjustment of focal plane position) mounted on a precision multi-axis alignment stage. The free-space coupling arrangement is illustrated in \autoref{fig1}.

\section{Laser-Induced Damage Threshold}
In order to assess the fiber’s damage threshold, laser power was progressively increased and monitored. All measurements in this section were performed using the 20 m fiber, at 1 and 30 kHz. Peak power densities were calculated for M$^2$=1.0 and M$^2$=1.2 (where M$^2$: beam quality factor), bracketing the manufacturer-specified beam quality range. The peak power density (I) at the fiber input facet was obtained (Table 2) by dividing the peak power by the effective beam area \cite{saleh2019fundamentals}, defined for a collimated Gaussian beam of spot diameter d at focus as: 
\begin{equation}
    d=\frac{4M^2 \lambda f}{\pi D} 
\end{equation}

\begin{equation}
    I=2\frac{P_{peak}}{A}=2\frac{P_{peak}}{\pi (d/2)^2} 
\end{equation}

where D is beam diameter before coupler and f is coupler focal length.
For a Gaussian beam profile, the peak power density is a factor of 2 higher than the average density over the beam area of a circular beam of diameter d ($I_{cir}=P_{peak}/A$). The $1/{e^2}$ beam diameter convention is typically used in this calculation. The end cap beam diameter ($d_z$) is calculated by back-propagating from the fiber core (where the focused waist is located) outward through the 400 $\mu$m of fused silica to the external facet. The key is the Rayleigh range \cite{pedrotti2018introduction}:

\begin{equation}
    z_R=\frac{n \pi w_o^2}{M^2 \lambda}
\end{equation}

\begin{equation}
    d_z=d \sqrt {1+(\frac{z}{z_R})^2}
\end{equation}

where n=1.46 is the refractive index for silica, $w_0$ is the beam waist at focal plane and z is the end cap length.
No instantaneous damage was observed at any tested peak power density (Table 2). Long-term resilience results are shown in Table 3.

\begin{table}[htbp]
\centering
\caption{Peak Power Density at Fiber Input  }
\label{tab:laser_parameters}
\begin{tabular}{|l|l|l|l|l|l|l|}
\toprule
\textbf{PRR} & \textbf{Beam} & \textbf{Peak} & \textbf{Beam} & \textbf{Peak} & \textbf{Beam} & \textbf{Peak} \\
\textbf{(kHz)} & \textbf{Quality} & \textbf{Power} & \textbf{Diameter ($d_z$)} & \textbf{Power} & \textbf{Diameter ($d$)} & \textbf{Power} \\
 & \textbf{Factor} & (W) & \textbf{(at end cap} & \textbf{Density} & \textbf{(at fiber} & \textbf{Density} \\
 & & & \textbf{facet)} ($\mu$m) & \textbf{(at end} & \textbf{core)} ($\mu$m) & \textbf{(at fiber} \\
 & & & & \textbf{cap facet)} & & \textbf{core)} \\
 & & & & (GW/cm$^2$) & & (GW/cm$^2$) \\
\midrule
1  & M$^2$=1.0 & 1323.53 & 53.70 & 0.117 & 3.46 & 28.2 \\
1  & M$^2$=1.2 & 1323.53 & 53.75 & 0.116 & 4.16 & 19.5 \\
30 & M$^2$=1.0 & 1646.71 & 36.22 & 0.320 & 5.17 & 15.7 \\
30 & M$^2$=1.2 & 1646.71 & 36.41 & 0.317 & 6.21 & 10.9 \\
\bottomrule
\end{tabular}
\end{table}

\begin{table}[htbp]
\centering
\caption{Fiber Resilience Under Extended Operation}
\label{tab:fiber_resilience}
\begin{tabular}{|l|l|l|l|l|}
\hline
\textbf{PRR (\textit{kHz})} & \textbf{Peak Power} & \textbf{Peak Power} & \textbf{\textit{Short-Term}} & \textbf{\textit{Long-Term}} \\
 & \textbf{Density (at end} & \textbf{Density (at fiber} & ($<1$~h) & ($>5$~h) \\
 & \textbf{cap facet)} & \textbf{core)} (mean value) & & \\
 & (mean value) & & & \\
 & (GW/cm$^2$) & (GW/cm$^2$) & & \\
\hline
1  & $\sim$0.12 & $\sim$24 & YES & NO \\ \hline
30 & $\sim$0.32 & $\sim$13 & YES & YES \\ \hline
\end{tabular}
\end{table}

As a result, the end cap was found to reduce peak power density by a factor of $\sim200$× at 1 kHz and $\sim40$× at 30 kHz at the air-silica interface, the most damage-susceptible surface, where any contamination, surface roughness or micro-defect can cause localized absorption and catastrophic failure under intense pulsed illumination. The protective role of end cap is more intense at 1 kHz due to the tighter focused spot from the larger input beam.

\section{Coupling Efficiency Measurements}

\begin{figure}[!ht]
 \includegraphics[width=0.8\textwidth]{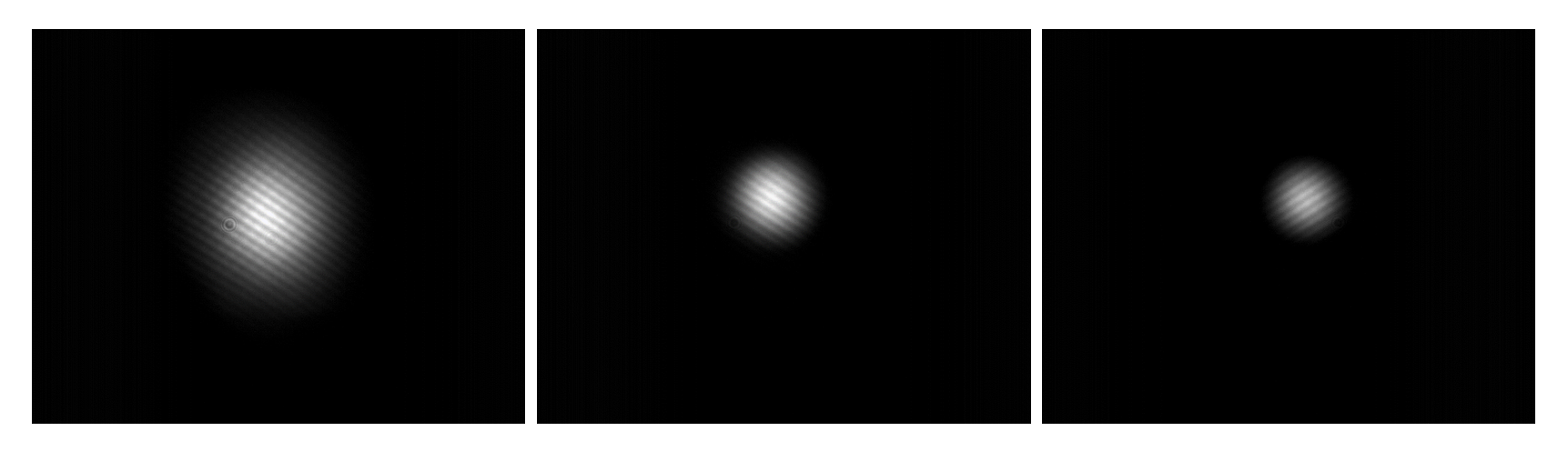}{}
	\centering
	\caption{Beam profiles at 1 kHz (left), 30 kHz (middle) and 40 kHz (right). The reduced beam diameter at 30 and 40 kHz is consistent with thermal lensing in the laser cavity.}
	\label{fig2}
\end{figure}

During the damage threshold measurements, the input beam diameter was observed to vary with repetition rate, an effect attributed to thermal lensing within the laser cavity under sustained high-rate operation\cite{liu2008effect,lu2020thermal}. This beam diameter variation was identified as a likely factor limiting coupling efficiency, since a mismatch between the incident beam at focal plane and the fiber mode field directly reduces spatial overlap at the input facet. In order to characterize the spatial beam profile at tested repetition rates, the output of the laser cavity was imaged using a FLIR beam profiling camera. The acquired images (\autoref{fig2}) revealed a clear, repetition-rate-dependent change in beam morphology: at 1 kHz the beam exhibited a notably larger and slightly elliptical intensity distribution, whereas at 30 and 40 kHz the profile became significantly more compact and nearly circular.

\begin{figure}[!ht]
 \includegraphics[width=0.4\textwidth]{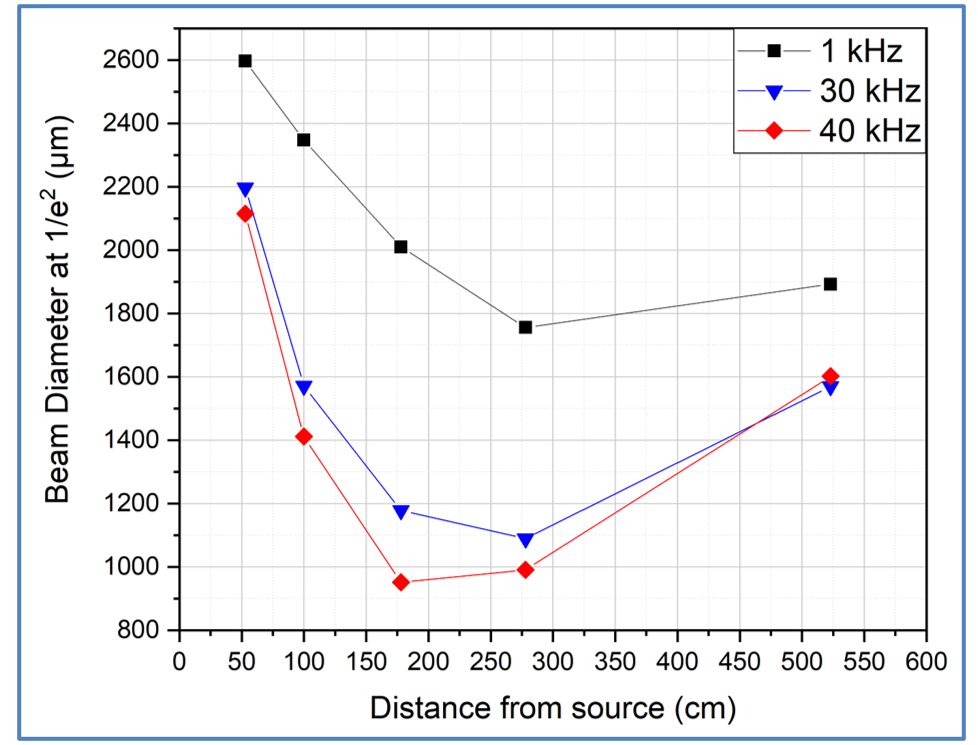}{}
	\centering
	\caption{Beam diameter behavior as a function of distance from source.}
	\label{fig3}
\end{figure}

\autoref{fig3} shows the beam diameter at the $1/e^2$ intensity level as a function of distance from the laser source, for repetition rates of 1 kHz, 30 kHz and 40 kHz. All three profiles exhibit a converging-then-diverging behavior along the propagation axis, consistent with the focusing action of the intracavity thermal lens. Notably, the minimum beam diameter, reached at approximately 270 cm from the source, decreases significantly with increasing repetition rate, from $\sim1800 \mu$m at 1 kHz, to $\sim1100 \mu$m at 30 kHz, and to $\sim950 \mu$m at 40 kHz, reflecting the progressively stronger thermal lensing effect at higher average powers-average laser output power also rises along with repetition rate. In our fiber coupling arrangement, the fiber coupler was placed at a position 100 cm from source. Accordingly, all beam diameter values used in the calculations were taken from \autoref{fig3} at the 100 cm position.

\begin{figure}[!ht]
 \includegraphics[width=0.4\textwidth]{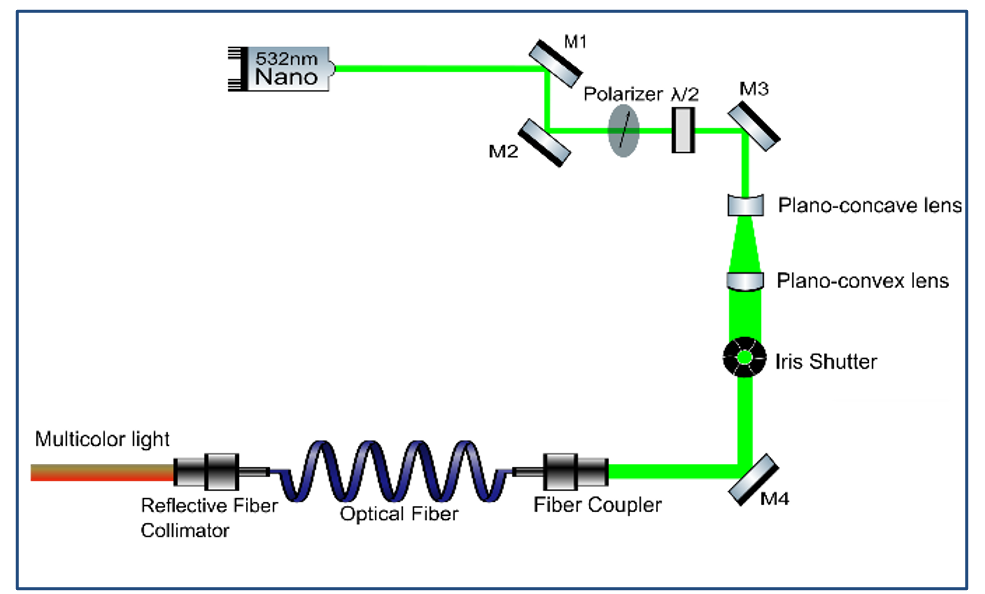}{}
	\centering
	\caption{Updated free-space optical arrangement incorporating a two-lens telescope and a variable iris shutter for beam diameter control, enabling optimized coupling efficiency at higher repetition rates.}
	\label{fig4}
\end{figure}

This repetition-rate-dependent change in beam size is consistent with thermal effects building up within the laser cavity at higher average powers which alter the beam propagation conditions and consequently the output beam diameter of the laser cavity. To address this, a two-lens telescope was introduced into the free-space beam path to expand the beam to a diameter large enough to create the flexibility to define the desired beam size at the fiber input. A variable iris shutter placed at the telescope output was then used to select the target beam diameter for each operating condition, ensuring optimal mode matching at the fiber coupler. The resulting optical arrangement is illustrated in \autoref{fig4}. An optical fiber of 5 m long, also provided by SQS FiberOptics, was used to conduct coupling efficiency measurements. As a direct result of this combined approach, the beam diameter was brought into optimal correspondence with the fiber mode field diameter, maximizing the spatial overlap at the input facet and yielding significantly improved coupling efficiency at high repetition rates. 

Coupling efficiency $\eta=P_{out}/P_{in}$ was measured under optimized alignment at all repetition rates. The average power at both the input and output of the fiber was measured using the same power meter as in Section 3, allowing direct determination of the coupling efficiency. Following the introduction of the telescope and iris shutter into the free-space coupling array, coupling efficiencies exceeding $45\%$ were achieved across all tested high repetition rates, including 40 kHz - a condition that had yielded particularly poor results due to a degraded beam profile unsuitable for efficient mode matching at the fiber coupler. The improvement confirms that the beam conditioning stage effectively compensated for the repetition-rate-dependent beam deformation, restoring the spatial overlap at the fiber input facet to achieve high coupling efficiency.

\begin{table}[htbp]
\centering
\caption{Measured Coupling Efficiencies}
\label{tab:coupling_efficiencies}
\begin{tabular}{|l|l|}
\hline
\textbf{\textit{Repetition Rate}} & \textbf{\textit{Maximum Coupling Efficiency}} \\ \hline
1 kHz                             & 53\%                                      \\ \hline
30 kHz                            & 51\%                                      \\ \hline
40 kHz                            & 46\%                                      \\ \hline
\end{tabular}
\end{table}

\section{Discussion}
\subsection{Elevated Damage Resistance}
Peak power densities during testing exceed typical published LIDT values $(\sim1 GW/cm^2)$ by manufacturers for uncoated silica fiber facets by nearly an order of magnitude. The absence of immediate damage strongly suggests the integrated end cap plays a decisive protective role, distributing optical intensity over a larger cross-sectional area before reaching the fiber core. Long-term degradation at 1 kHz likely reflects cumulative thermal effects at the highest achievable power densities, rather than catastrophic failure.

\subsection{Beam Diameter Management as a Key Enabler for Array Performance}
The identification of thermally induced beam diameter variation as a primary limiting factor, and its successful mitigation through telescope-based beam conditioning and iris aperture control, represents a significant step forward in the optimization of the SRS fiber array. The recovery of coupling efficiency exceeding $45\%$ at 30 and 40 kHz -achieved solely by correcting the spatial mode mismatch - demonstrates that coupling performance can be further improved by optimizing the beam delivery optics alone, without any changes to the fiber or laser source. Equally important is the robustness of the PM 460HP end-capped fiber patchcords, which sustained operation under extreme peak power densities without instantaneous damage and maintained integrity over extended durations. 
\subsection{Above-Specification Coupling Efficiency}
Coupling efficiencies of $53\%, 51\%$ and $46\%$ at 1 kHz, 30 kHz and 40 kHz respectively, substantially exceed the manufacturer’s typical value of $\sim30\%$. The slight variation in efficiency across repetition rates is attributed to experimental limitations rather than fundamental differences in coupling performance, as the optical alignment and beam conditioning were independently optimized for each operating condition.

\section{Conclusions}
Key findings:\\
•	Damage threshold: No instantaneous damage at peak power densities exceeding $10^{10} W/cm^2$; the integrated end cap identified as the principal protective mechanism.\\
•	Long-term resilience: Sustained at 30 kHz (>5 h) $(\sim13 GW/cm^2$ at fiber core); gradual degradation at 1 kHz $(\sim 24 GW/cm^2$ at fiber core) under prolonged high-intensity exposure.\\
•	Beam-profile dependence: Repetition-rate-dependent beam profile and output power changes are the primary driver of coupling efficiency variation.\\
•	Beam diameter management: Beam output of the laser cavity conditioning via a telescope and iris shutter recovered efficiencies exceeding $45\%$ for all experimental conditions, validating beam diameter control as a critical optimization step for the SRS fiber array.

\section{Acknowledgments}This work was supported by the H2020 FETOPEN project “DynAMic” (grant number: EC-GA-863203), the NSRF 2014–2020 “BIOIMAGING-GR” (grant number: MIS 5002755), and EIC Pathfinder Open 2021 project "SWOPT" (grant number: 101046667).

\bibliographystyle{unsrtnat}
\bibliography{references}  

@article{abbasi2021highly,
  title={Highly flexible fiber delivery of a high peak power nanosecond Nd:YAG laser beam for flexiscopic applications},
  author={Abbasi, Hamed and Canbaz, Fatih and Guzman, Raphael and Cattin, Philippe C and Zam, Azhar},
  journal={Biomedical Optics Express},
  volume={12},
  number={1},
  pages={444--458},
  year={2021},
  publisher={Optica Publishing Group},
  doi={10.1364/boe.405825},
  url={https://doi.org/10.1364/boe.405825}
}

@article{hunter1996understanding,
  title={Understanding high-power fiber-optic laser beam delivery},
  author={Hunter, B V and Leong, K H and Miller, C B and Golden, J F and Glesias, R D and Laverty, P J},
  journal={Journal of Laser Applications},
  volume={8},
  number={6},
  pages={307--316},
  year={1996},
  publisher={Laser Institute of America},
  doi={10.2351/1.4745437},
  url={https://doi.org/10.2351/1.4745437}
}

@book{saleh2019fundamentals,
  title={Fundamentals of Photonics},
  author={Saleh, Bahaa E A and Teich, Malvin Carl},
  edition={3rd},
  year={2019},
  publisher={John Wiley \& Sons},
  isbn={9781119506874}
}

@book{pedrotti2018introduction,
  title={Introduction to Optics},
  author={Pedrotti, Frank L and Pedrotti, Leno M and Pedrotti, Leno S},
  edition={Third edition},
  year={2018},
  publisher={Cambridge University Press},
  doi={10.1017/9781108552493},
  url={https://doi.org/10.1017/9781108552493}
}

@article{liu2008effect,
  title={Effect of thermal lens on beam quality and mode matching in LD pumped Er--Yb-codoped phosphate glass microchip laser},
  author={Liu, S and Song, F and Cai, H and Li, T and Tian, B and Wu, Z and Tian, J},
  journal={Journal of Physics D: Applied Physics},
  volume={41},
  number={3},
  pages={035104},
  year={2008},
  publisher={IOP Publishing},
  doi={10.1088/0022-3727/41/3/035104},
  url={https://doi.org/10.1088/0022-3727/41/3/035104}
}

@article{lu2020thermal,
  title={Thermal focal length determination of a laser crystal by modulating the pump source},
  author={Lu, H and Mangaiyarkarasi, D},
  journal={Laser Physics},
  volume={30},
  number={6},
  pages={065001},
  year={2020},
  publisher={IOP Publishing},
  doi={10.1088/1555-6611/ab8af3},
  url={https://doi.org/10.1088/1555-6611/ab8af3}
}






\end{document}